\documentclass[aps, prd, superscriptaddress, nofootinbib, preprintnumbers]{revtex4}

\usepackage{amsmath,amssymb,bm}
\usepackage[dvipdfmx]{graphicx}
\usepackage{color}
\usepackage{float}
\usepackage{braket}

\renewcommand\d{\partial}

\allowdisplaybreaks

\begin{document}

\title{Vector Meson Masses from Hidden Local Symmetry in Constant Magnetic Field}
\author{Mamiya Kawaguchi\footnote{mkawaguchi@hken.phys.nagoya-u.ac.jp}}
      \affiliation{ Department of Physics, Nagoya University, Nagoya 464-8602, Japan.} 
\author{Shinya Matsuzaki\footnote{synya@hken.phys.nagoya-u.ac.jp}}
      \affiliation{ Department of Physics, Nagoya University, Nagoya 464-8602, Japan.}       
\affiliation{ Institute for Advanced Research, Nagoya University, Nagoya 464-8602, Japan.}

\date{\today}

\def\theequation{\thesection.\arabic{equation}}
\makeatother

 \begin{abstract}
We discuss the magnetic responses of vector meson masses 
based on the hidden local symmetry (HLS) model in constant magnetic field,  
described by the lightest two-flavor system including 
the pion, rho and omega mesons in the spectrum. 
The effective masses influenced under the magnetic field are 
evaluated in a way of 
the derivative/chiral expansion established in the HLS model. 
At the leading order ${\cal O}(p^2)$ the g-factor of the charged rho meson is 
fixed to be 2, implying that the rho meson at this order is treated just like a point-like spin-1 particle.  
Beyond the leading order,   
one finds anomalous magnetic interactions of the charged rho meson, 
involving the anomalous magnetic moment, 
which give corrections to the effective mass.  
It is then suggested that up to ${\cal O}(p^4)$ 
the charged rho meson tends to become massless.  
Of interest is that nontrivial magnetic-dependence of neutral mesons emerges 
to give rise to the significant mixing among neutral mesons. 
Consequently, it leads to the dramatic enhancement of the omega meson mass, 
which is testable in future lattice simulations. 
Corrections from terms beyond ${\cal O}(p^4)$ are also addressed.   

\end{abstract}
\maketitle
\section{Introduction}

Exploring the quantum chromodynamics (QCD) in 
external magnetic field has recently attracted a lot of interest, 
such as the presumable presence of the strong magnetic field in the neutron star or magnetar, 
an early stage of heavy ion collisions, 
and some topics related to physics on the early Universe.  
For instance, in off-central heavy ion collisions,  
the scale of the magnetic field reaches up to about a few hundreds of MeV, 
which could also be related to dynamics of quark-gluon plasma.  
In this respect, several fascinating QCD phenomena in the magnetic field 
have been proposed:  
the chiral magnetic effect, 
the magnetic catalysis or inverse magnetic catalysis, and so forth. 

More on striking and exotic magnetic phenomena involves 
hadron physics: some of studies implies that vector meson (rho meson) can condense 
due to the presence of a strong magnetic field, 
which is naively expected from the Landau-quantized mass of charged particles with spin 1.  
In this respect, several works have been done based on effective models for 
QCD~\cite{Chernodub:2010qx,Callebaut:2011uc,Ammon:2011je,Frasca:2013kka,Liu:2014uwa,Filip:2015mca}, 
and also some objection against the rho meson condensation from current lattice simulation 
has been reported, say, in Ref.~\cite{Hidaka:2012mz}.    
Understanding the QCD in magnetic field  is  therefore getting excited   
not only on the field theoretical ground, but also on some phenomenological aspect 
involving other research fields.

In this paper, we discuss the magnetic responses of vector meson masses 
based on a chiral effective model including vector mesons as 
gauge bosons of some gauge symmetry: it is so called   
the hidden local symmetry (HLS) model~\cite{Bando:1984ej,Bando:1987br}. 
Setting the background photon gauge to be a constant magnetic field, 
and employing the lightest two-flavor system including 
the pion, rho and omega mesons in the spectrum, 
we evaluate the effective masses influenced under the magnetic field  
in a way of the derivative/chiral expansion established in the HLS model.

At the leading order, the g-factor of the charged rho meson is 
fixed to be 2, implying that the rho meson at this order is dealt with, just like a point-like spin-1 particle. 
Going beyond the leading order,  
we find anomalous magnetic interactions of the charged rho meson.  
The anomalous magnetic moment arises from ${\cal O}(p^4)$ terms, 
and its magnitude can be manifestly controlled in the derivative/chiral 
expansion.  
It turns out that up to ${\cal O}(p^4)$ the charged rho meson tends to become 
massless.  
More remarkably, nontrivial magnetic-dependence of neutral mesons emerges 
from ${\cal O}(p^4)$ terms of the HLS model. 
This is tied with the significant mixing among neutral mesons under the magnetic field, 
breaking the spin and Lorentz invariance. 
As a consequence, we observe the dramatic enhancement of the omega meson mass, 
which is testable in future lattice simulations.

This paper is organized as follows: 
in Sec.~\ref{HLS-model} 
we make a brief review of the HLS model  
and formulate the model in a constant magnetic field. 
The constant magnetic effect on vector meson masses 
are then evaluated in Sec.~\ref{effective-masses} 
including terms up to  ${\cal O}(p^4)$.  
In Sec.~\ref{summary},  
summary of this paper is given and 
we make comments on possible corrections from terms higher than ${\cal O}(p^4)$ 
in the derivative/chiral expansion. 
The details of the calculation on the Landau quantization 
for the vector meson mass are presented in Appendix~\ref{Landau}.

\section{The HLS Model in Constant Magnetic Field} 
\label{HLS-model}

In this section we start with a brief review of the HLS model~\cite{Bando:1984ej,Bando:1987br} 
described by the pion, $\rho$ and $\omega$ mesons together with the photon,  
and then formulate the model in a constant magnetic field.

\subsection{Review of Hidden Local Symmetry Model}

The HLS formalism is a simple extension from the nonlinear realization of the chiral 
symmetry. 
To see how it works, 
we first write the chiral field $U$ in the nonlinear realization,  
$ 
U =e^{i\pi^i \tau^i/F_\pi}   
$,  
where $\pi^i$ ($i=1,2,3$) are pion fields, 
$\tau^i$ are Pauli matrices, 
and $F_{\pi}$ is the decay constant  
associated with the spontaneous breaking of the global chiral symmetry,  
$G=SU(2)_L\times SU(2)_R \times U(1)_V$ $\to$ 
$H=U(2)_{V=L+R} \times U(1)_V$.  
The chiral field $U$ transforms under the $G$ 
as 
$ 
U\rightarrow g_L \cdot U \cdot g_R^\dagger 
$, 
with $g_{L,R}\in G$. 
One then should note an arbitrariness or a gauge degree of freedom (HLS) 
in dividing the $U$ into a product of nonlinear bases, 
$\xi_L^\dagger$ and $\xi_R$, in such a way that they transform as 
$    
\xi_{L,R} 
\rightarrow h(x) \cdot \xi_{L,R} \cdot g_{L,R}^\dagger 
$, 
where $h(x) \in H_{\rm local}=[U(2)_V]_{\rm local}$ and $g_{L,R} \in G_{\rm global}=SU(2)_L \times SU(2)_R \times U(1)_V$.  
Thus, introducing the redundant (spontaneously broken) gauge symmetry $H_{\rm local}$ (HLS),  
 the chiral system can always be extended from the coset space $G/H=[SU(2)_L\times SU(2)_R \times U(1)_V]/[SU(2)_V \times U(1)_V]$  
to $G_{\rm global} \times H_{\rm local}/H_{\rm diag} = [SU(2)_L \times SU(2)_R \times U(1)_V \times [U(2)_V]_{\rm local}]/
[SU(2)_{V'} \times U(1)_V]$, where 
$H_{\rm diag}=SU(2)_{V'}$ denotes the diagonal sum of 
$H_{\rm global} \in G_{\rm global}$ and $H_{\rm local}$.

In association with the HLS, 
the gauge fields ($V_\mu$)  
are introduced, 
which  transform under the HLS as 
$ 
V_{\mu} \rightarrow h(x) \cdot V_\mu \cdot 
h^\dagger(x) + ih(x)\d_\mu h^\dagger(x)
$. 
The $\rho$ and $\omega$ meson fields are embedded in 
these HLS gauge fields as 
\begin{equation} 
 V_\mu = g  \rho_\mu + g' \omega_\mu 
\,,\qquad 
\rho_\mu = \rho_\mu^i \frac{\tau^i}{2} \quad (i=1,2,3) 
\,, \qquad 
\omega_\mu = \omega_\mu \cdot \frac{\tau^0}{2}  
\quad (\tau^0 \equiv {\bf 1}_{2 \times 2})  
\,, \label{embedding}
\end{equation}
where $g$ and $g'$ stand for the HLS gauge couplings corresponding to 
the $SU(2)$ and $U(1)$ parts in $H_{\rm local} = [U(2)_V]_{\rm local}$, respectively. 
These vector mesons get massive by eating the would-be (fictitious) Nambu-Goldstone (NG) bosons 
${\cal P}={\cal P}^i \tau^i/2 + {\cal P}^0 \tau^0/2$ (just like the Higgs mechanism) 
embedded in the $\xi_{L,R}$ as 
\begin{eqnarray} 
  \xi_{L,R}=e^{i {\cal P}^i\frac{\tau^i}{2}/F_{\cal P}} \cdot 
e^{i {\cal P}^0 \frac{\tau^0}{2}/F_{{\cal P}^0}} \cdot 
  e ^{\mp i \pi^i \frac{\tau^i}{2}/F_\pi} 
\,, 
  \end{eqnarray}
where 
$F_{{\cal P}^{(0)}}$ are the associated decay constant giving the scale of the 
vector meson masses. 
After gauge fixing of $H_{{\rm local}}$, say, by the unitary gauge ${\cal P} \equiv 0$, 
the $H_{\rm diag}$ becomes $H$ of the usual nonlinear sigma model manifold $G/H$~\cite{Bando:1984ej,Bando:1987br}.

 To construct the Lagrangain invariant under the $G_{\rm global} \times H_{\rm local}$, 
it is convenient to introduce Maurer-Cartan 1-forms: 
 \begin{eqnarray}
 \hat{\alpha}_{\perp,|| \mu} 
=\frac{1}{2i}(D_{\mu}\xi_R\cdot \xi^{\dagger}_R \mp D_{\mu}\xi_L\cdot\xi^{\dagger}_L) 
\,, 
\label{1-forms}
 \end{eqnarray}
where 
\begin{eqnarray}
D_{\mu}\xi_{L,R}=\partial_{\mu}\xi_{L,R}-iV_{\mu}\xi_{L,R} 
\,. \label{cov:1}
\end{eqnarray} 
One then finds that 
the 1-forms in Eq.(\ref{1-forms}) 
transform homogeneously under the $G_{\rm global} \times H_{\rm local}$ as 
$\hat{\alpha}_{\perp, || \mu } \to h(x) \cdot \hat{\alpha}_{\perp, || \mu} \cdot h^\dag(x)$. 
In addition to the 1-forms, the field strength of the HLS gauge fields $V_\mu$ 
is introduced as 
\begin{equation} 
V_{\mu\nu}=\d_\mu V_\nu-\d_\nu V_\mu-i[V_\mu,V_\nu]
\,, 
\end{equation} 
which transform homogeneously in the same way as the 1-forms.

With these building blocks at hand, one can readily write down 
the Lagrangian invariant under the $G_{\rm global} \times H_{\rm local}$ 
(and charge and parity conjugations). 
At the leading order of derivative expansion (${\cal O}(p^2)$), 
the  Lagrangian goes like  
\begin{eqnarray}
{\cal L}_{(2)}
&=&
F_\pi^2{\rm tr}[\hat\alpha_{\perp\mu}\hat\alpha_\perp^\mu]\nonumber\\
&&+
\frac{m_\rho^2}{g^2}{\rm tr}[\hat\alpha_{\parallel\mu}\hat\alpha_\parallel^\mu]
+\bigl(\frac{m_\omega^2}{2g'^2}-\frac{m_\rho^2}{2g^2}\bigl){\rm tr}[\hat\alpha_{\parallel\mu}]{\rm tr}[\hat\alpha_\parallel^\mu]\nonumber\\
&&-\frac{1}{2g^2}{\rm tr}[V_{\mu\nu}V^{\mu\nu}]
-\bigl(\frac{1}{4g'^2}-\frac{1}{4g^2}\bigl){\rm tr}[V_{\mu\nu}]{\rm tr}[V^{\mu\nu}]
\,, 
\label{1-formLagrangian}
\end{eqnarray}
where 
we have assigned the derivative order for the parameters in 
the vector meson sector as 
\begin{equation} 
g \sim g' \sim m_\rho \sim m_\omega \sim O(p)
\,, 
\label{order-counting}
\end{equation} 
which makes it possible to perform the systematic expansion 
in terms of the chiral perturbation theory including the HLS~\cite{Tanabashi:1993np,Harada:2003jx}.  
Instead of the decay constants $F_{{\cal P}^{(0)}}$ for the HLS, 
in Eq.(\ref{1-formLagrangian}) we have used $m_\rho$ and $m_\omega$ taking into account the form of the embedding in Eq.(\ref{embedding}).

Independently of the HLS, 
one can freely gauge the $G_{{\rm global}}$ 
by introducing the external gauge fields ${\cal L}_\mu$ and ${\cal R}_\mu$ 
including the photon field $A_\mu$, 
as 
${\cal L}_\mu = {\cal R}_\mu = e Q_{\rm em} A_\mu$, where 
$e$ is the electromagnetic coupling and 
$Q_{\rm em}=\tau_3/2+ \tau^0/6 = {\rm diag}(2/3, -1/3)$.  
Then the covariant derivatives in Eq.(\ref{cov:1}) are changed as  
\begin{eqnarray} 
D_{\mu}\xi_{L,R} \to 
D_{\mu}\xi_{L,R} =\partial_{\mu}\xi_{L,R}-iV_{\mu}\xi_{L,R}+i\xi_{L,R} \, eQ_{\rm em} A_\mu 
\,,  
\end{eqnarray}
as well as the 1-forms in Eq.(\ref{cov:1}).

The pion mass term can be incorporated through introduction of a spurion field $\hat{\chi}=\xi_L \chi \xi_R^\dag$, where $\chi$ transforms as in the same way as the chiral field $U$ 
and is counted as ${\cal O}(p^2)$ in the derivative/chiral expansion.    
The mass-term Lagrangian at the leading order  
is then written in a manner invariant under the 
chiral symmetry and HLS~\cite{Harada:2003jx}: 
\begin{equation} 
 {\cal L}_{(2)}^\chi = \frac{F_\pi^2}{4} {\rm tr}[\hat{\chi}^\dag + \hat{\chi}] 
\,. \label{chi:term}
\end{equation} 
 When the spurion field $\chi$ gets the vacuum expectation value, $\langle \chi \rangle 
= m_\pi^2 \cdot {\bf 1}_{2 \times 2}$ (assuming the isospin symmetric form), 
taking the unitary gauge of the HLS (${\cal P}\equiv 0$) and 
expanding the 1-forms in powers of the pion fields, 
\begin{eqnarray}
 \hat{\alpha}_{\perp \mu}&=& 
\frac{1}{F_\pi}\d_\mu\pi-\frac{i}{F_\pi}[e Q_{\rm em} A_\mu ,\pi]+\cdots 
\nonumber \\
 \hat{\alpha}_{\parallel \mu}&=& 
-V_\mu+ e Q_{\rm em} A_\mu + \cdots 
\,, \label{expand:alpha}
\end{eqnarray}
one finds that the pions get the mass through the Lagrangian ${\cal L}_{(2)}^\chi$ in Eq.(\ref{chi:term}).

Including the pion mass $m_\pi \sim {\cal O}(p)$ in addition to 
the vector meson masses counted as in Eq.(\ref{order-counting}), 
one can thus systematically discuss the  
the phenomenology of the vector mesons and pions coupled to the 
photon 
in a way of the derivative/chiral expansion as in the literature~\cite{Tanabashi:1993np,Harada:2003jx}  
with the expansion coefficients, 
\begin{equation} 
\frac{p}{(4\pi F_\pi)} \sim 
 \frac{m_\pi}{(4 \pi F_\pi)} \sim 
\frac{m_{\rho,\omega}}{(4 \pi F_\pi)} \sim {\cal O}(p)
\,.\label{expansion-parameters}
\end{equation}

\subsection{Expanding in Constant Magnetic Field}\label{ecm}

Now we formulate the HLS model in a constant magnetic field 
based on the Lagrangian Eq.(\ref{1-formLagrangian}). 
To this end, using Eqs.(\ref{embedding}) 
and (\ref{expand:alpha}), 
we shall first expand Eq.(\ref{1-formLagrangian}) 
in terms of the NG boson fields, 
and focus on the vector meson part coupled to the external photon field,  
to find 
\begin{eqnarray}
{\cal L}_{\rho,\omega}
&=&
\frac{1}{2}\omega_\mu(g^{\mu\nu}\d^2-\d^\mu\d^\nu)\omega_\nu
+\rho^+_\mu(g^{\mu\nu}\d^2-\d^\mu\d^\nu)\rho^-_\nu
 +\frac{1}{2}\rho^0_\mu(g^{\mu\nu}\d^2-\d^\mu\d^\nu)\rho^0_\nu
\nonumber \\  
&& 
-ig\rho^{0\nu}(\partial_\mu\rho_\nu^-\rho^{+\mu}-\partial_\mu\rho^+_\nu\rho^{-\mu})
-ig\rho^{0\mu}(\partial_\mu\rho_\nu^+\rho^{-\nu}-\partial_\mu\rho^-_\nu\rho^{+\nu})
-ig\partial_\mu\rho^0_\nu(\rho^{-\mu}\rho^{+\nu}-\rho^{+\mu}\rho^{-\nu}) 
\nonumber \\ 
&& 
- V_\omega - V_\rho
\label{omegarhoLg}
\,, 
\end{eqnarray}
where $\rho^{\pm}=(\rho^1\mp i\rho^2)/\sqrt2$, $\rho^0 \equiv \rho^3$ 
and 
\begin{eqnarray}
V_\omega 
&=&
-\frac{m_\omega^2}{2}\Bigl(\omega_\mu-\frac{e}{3g'}A_\mu\Bigl)^2
\,, \nonumber \\
V_\rho&=&
-\frac{m_\rho^2}{2}\Bigl( \rho_\mu^0-\frac{e}{g}A_\mu\Bigl)^2
+g^2(\rho^0_\mu\rho^{0\mu})(\rho_\nu^+\rho^{-\nu})-g^2\rho^0_{\mu}\rho^{0\nu}\rho^+_\nu\rho^{-\mu}
\nonumber \\
&& 
-m_\rho^2\rho_\mu^+\rho^{-\mu}
+\frac{g^2}{2}(\rho_\mu^+\rho^{-\mu})^2-\frac{g^2}{2}(\rho_\nu^+\rho^{+\nu})(\rho_\mu^-\rho^{-\mu})
\,. 
\label{vpot}
\end{eqnarray}
Note that the potential terms $V_\omega$ and $V_\rho$ contain the mixing between the neutral vector 
mesons and the photon.

We now consider a constant magnetic field $B$ 
oriented to the $z$-direction in the four-dimensional space-time. 
It can be observed by acquiring the vacuum expectation value of the photon field 
like 
\begin{eqnarray}
\langle A_\mu \rangle 
=(0,-By/2,Bx/2,0)
\,, 
\label{symmetry}
\end{eqnarray}
where we took the symmetric gauge. 
Then the potentials in Eq.(\ref{vpot}) implies 
shifts of neutral vector potentials in $V_\omega$ and $V_\rho$ by nonlocal 
vacuum expectation value of the photon field  $\braket{A_\mu}$(constant $B$).
Thus, due to the presence of the constant $B$, 
the stationary point for the $V_\omega$ and $V_\rho$ is changed 
from $\langle \rho_\mu^{\pm, 0} \rangle = \langle \omega_\mu \rangle =0$ to  
\begin{eqnarray}
\braket{\rho^{0\mu}} 
=\frac{e}{g}\braket{A^\mu}
\,, \qquad 
\braket{\omega^\mu}=\frac{e}{3g'}\braket{A^\mu}
\,, \qquad 
\braket{\rho^{\pm\mu}}=0 
\,. 
\end{eqnarray}
Expanding the fields around these vacuum expectation values 
as 
\begin{eqnarray} 
\rho^{0\mu}&=&\braket{\rho^{0\mu}}+\tilde{\rho}^{0\mu}\;\;\;\;
\omega^\mu=\braket{\omega^\mu}+\tilde{\omega}^\mu 
\,, \nonumber \\
\rho^{+\mu}&=&\tilde{\rho}^{+\mu}\;\;\;\;\;\;\;\;\;\;\;\;\;\;\;A^\mu=\braket{A^\mu}
\,. 
\end{eqnarray}
we see that the Lagrangian Eq.(\ref{omegarhoLg}) is modified to be 
\begin{eqnarray}
{\cal{L}}_{\rho,\omega}&&=
-\frac{1}{2} 
(D_\mu\tilde{\rho}^-_\nu - D_\nu\tilde{\rho}^-_\mu) 
 (D^\mu\tilde{\rho}^{+\nu} - D^\nu\tilde{\rho}^{+\mu}) 
-\frac{i}{2} e \langle F_{\mu\nu} \rangle 
(\tilde{\rho}^{-\mu}\tilde{\rho}^{+\nu} - \tilde{\rho}^{+\mu}\tilde{\rho}^{-\nu}) 
\nonumber\\
&&
-\frac{1}{4} \left( \tilde{\rho}^0_{\mu\nu} - \frac{e}{g} \langle F_{\mu\nu} \rangle \right)^2 
-\frac{1}{4} \left( \tilde{\omega}^0_{\mu\nu} - \frac{e}{3 g'} \langle F_{\mu\nu} \rangle \right)^2 
\nonumber\\ 
&& 
+ 
i g \left[ 
\tilde{\rho}^{0 \mu} \tilde{\rho}^{+\nu} (D_\mu \tilde{\rho}_\nu^- - 
D_\nu \tilde{\rho}_\mu^- ) 
-  
\tilde{\rho}^{0 \mu} \tilde{\rho}^{-\nu} (D_\mu \tilde{\rho}_\nu^+ - 
D_\nu \tilde{\rho}_\mu^+ ) 
\right] 
- \frac{i}{2} g 
\tilde{\rho}_{\mu\nu}^0 
 (\tilde{\rho}^{-\mu}\tilde{\rho}^{+\nu} - \tilde{\rho}^{+\mu}\tilde{\rho}^{-\nu}) 
\nonumber \\ 
&&
+\frac{1}{2}m_\omega^2\tilde{\omega}_\mu\tilde{\omega}^\mu 
+ m_\rho^2\tilde{\rho}_\mu^+\tilde{\rho}^{-\mu}
+\frac{1}{2}m_\rho^2\tilde{\rho}_\mu^0\tilde{\rho}^{0\mu}  
\nonumber \\ 
&& 
- \frac{1}{2} g^2 \left[ 
( \tilde{\rho}_\mu^+ \tilde{\rho}^{- \mu})^2 
- ( \tilde{\rho}_\mu^+ \tilde{\rho}^-_\nu)^2 
+ (\tilde{\rho}_\mu^0)^2 \tilde{\rho}_\mu^+ \tilde{\rho}^{- \mu} 
- 
\tilde{\rho}_\mu^0 \tilde{\rho}_\nu^0 \tilde{\rho}^{+ \mu} \tilde{\rho}^{- \nu} 
\right] 
\,,  
\label{frhoLagrangian}
\end{eqnarray}
where 
\begin{eqnarray} 
D_\mu \tilde{\rho}^{\pm}_\nu 
&=&(\d_\mu\mp ie \langle A_\mu \rangle ) \tilde{\rho}^{\pm}_{\nu} 
\,, \nonumber \\ 
\langle F_{\mu\nu} \rangle 
&=& \partial_\mu \langle A_\nu \rangle - \partial_\nu \langle A_\mu \rangle
\,, \nonumber \\
\tilde{\rho}_{\mu\nu}^0 &=& 
\partial_\mu \tilde{\rho}_\nu^0 - \partial_\nu \tilde{\rho}_\mu^0 
\,, \nonumber \\ 
\tilde{\omega}_{\mu\nu} &=& 
\partial_\mu \tilde{\omega}_\nu - \partial_\nu \tilde{\omega}_\mu 
\,.  
\end{eqnarray}
The residual $U(1)_{\rm em}$ gauge invariance 
is manifest in the expression of Eq.(\ref{frhoLagrangian}). 
This should be so since the formulation addressed here is nothing 
but a sort of the background field method for the HLS gauge and photon 
fields, which surely ensures the 
gauge invariance for the background fields at the Lagrangian level.  
Note also from the last term of the first line in Eq.(\ref{frhoLagrangian}) 
that the HLS formalism fixes the magnetic moment of the charged $\rho$ to be 2, 
though it is in general arbitrary by only imposing the $U(1)_{\rm em}$ gauge invariance  
(More explicitly, see the later discussion).

\section{Effective Masses in Constant Magnetic Field from terms of ${\cal O}(eB)$}
\label{effective-masses}

In this section we discuss the vector meson masses influenced under the 
constant magnetic field based on the HLS model. 
We shall focus on the magnetic effect on the order of 
${\cal O}(eB)$ in the HLS model, which turns out to arise from not only terms of 
the leading order of ${\cal O}(p^2)$ in the Lagrangian Eq.(\ref{frhoLagrangian}), 
but also those of the next-to leading order of ${\cal O}(p^4)$ 
in the derivative/chiral expansion.

\subsection{Charged Vector Meson Mass }\label{transverse}

Looking at the constant magnetic field configuration in Eq.(\ref{symmetry}) 
one can find that only the charged $\rho$ mesons transversely polarized along 
the magnetic field $B$, $\tilde{\rho}^{\pm x,y}$, 
get affected. 
Taking into account the $B$-dependent mass term, 
which arises from the last term of the line one in Eq.(\ref{frhoLagrangian}), 
the $\tilde{\rho}^{\pm x,y}$ fields mix via 
\begin{eqnarray}
(\tilde{\rho}^{-x}, \tilde{\rho}^{-y})
\begin{pmatrix}
m_\rho^2&ieB\\
-ieB&m_\rho^2
\end{pmatrix}
\begin{pmatrix}
\tilde{\rho}^{+x}\\
\tilde{\rho}^{+y}
\end{pmatrix}
\,. \label{mass-basis}
\end{eqnarray}
To diagonalize this matrix we introduce 
\begin{eqnarray}
\Phi&=&\frac{1}{\sqrt{2}}(\tilde{\rho}^{+x}+i\tilde{\rho}^{+y})
\,, \nonumber \\
\phi&=&\frac{1}{\sqrt{2}}(\tilde{\rho}^{+x}-i\tilde{\rho}^{+y})
\,. 
\end{eqnarray}
The corresponding mass eigenvalues are given by 
\begin{eqnarray}
m_\Phi^2&=& m_\rho^2+eB 
\,, \nonumber \\
m_\phi^2&=& m_\rho^2-eB
\,. 
\end{eqnarray}
In terms of the mass-eigenstate fields in Eq.(\ref{mass-basis}),  
Eq.(\ref{frhoLagrangian}) is written as 
\begin{eqnarray}
{\cal{L}}_{\phi,\Phi}&=&-m_\rho^2(\Phi^*\Phi+\phi^*\phi)-eB\Phi^*\Phi+eB\phi^*\phi
+\d_t\Phi^*\d_t\Phi-\d_z\Phi^*\d_z\Phi 
+\d_t\phi^*\d_t\phi-\d_z\phi^*\d_z\phi 
\nonumber\\
&&-\frac{1}{2}D_x(\Phi^*-\phi^*)D_x(\Phi-\phi)
-\frac{1}{2}D_y(\Phi^*+\phi^*)D_y(\Phi+\phi)\nonumber\\
&&
-\frac{1}{2i}D_x(\Phi^*-\phi^*)D_y(\Phi+\phi)
+\frac{1}{2i}D_y(\Phi^*+\phi^*)D_x(\Phi-\phi)
\,, 
\label{phiLagrangian}
\end{eqnarray}
where terms of cubic order in the $\phi$ and $\Phi$ fields have been discarded. 
Solving the coupled equations of motion for $\phi$ and $\Phi$ 
one finds that the mass spectra are quantized   
by the Landau levels. 
We provide the detailed derivation in Appendix~\ref{Landau}. 
The effective mass in the lowest Landau level (LLL) is thus found to be  
\begin{eqnarray} 
m_\rho^2(eB)=m_\rho^2-eB({\cal G}-1) \;\;\;\;\;\; {\rm with}\;\;{\cal G}=2
\,, 
\label{G:op2}
\end{eqnarray}
where ${\cal G}$ is the g-factor of the charged $\rho$ meson. 
Remarkable to note is that the $\rho$ meson g-factor is unambiguously fixed to be 2 
in the HLS formalism 
as noted in the previous section.

The g-factor of the charged $\rho$ meson actually gets corrections, 
i.e., the anomalous magnetic moment, from 
terms of ${\cal O}(p^4)$ in the derivative/chiral expansion of the HLS model. 
  Examining the list of the ${\cal O}(p^4)$ terms in the literature~\cite{Tanabashi:1993np,Harada:2003jx} one can easily find 
that such corrections arise from the following operators: 
\begin{eqnarray} 
{\cal L}_{(4)}&=&z_3{\rm tr}[V_{\mu\nu}\hat{\cal V}^{\mu\nu}]
+iz_7{\rm tr}[\hat{\cal V}_{\mu\nu}\hat\alpha_{\parallel}^\mu\hat\alpha_{\parallel}^\nu]
\,, \label{z3z7}
\end{eqnarray} 
where  
\begin{equation} 
\hat{\cal V}_{\mu\nu}= 
\frac{1}{2} e \langle F_{\mu\nu}\rangle  
\left[ 
 \xi_L Q_{\rm em} \xi_L^\dag +  \xi_R Q_{\rm em} \xi_R^\dag
\right]  
\,.  
\end{equation} 
Including these $z_3$ and $z_7$ terms,  
one finds that 
the effective mass in Eq.(\ref{G:op2}) is now modified by 
the anomalous magnetic moment to be 
\begin{eqnarray}
m_\rho^2(eB)
&=&m_\rho^2-eB \left[ \left( {\cal G} -1\right) - g^2 z_3 +\frac{g^2 z_7}{2} \right] 
\,, \qquad  
{\rm with} \qquad {\cal G}=2 
 \nonumber \\ 
&=& 
\Bigg\{ m_\rho^2-eB  \Bigg\} 
+ \Bigg\{ \left(  z_3 - \frac{z_7}{2} \right) g^2 \, eB \Bigg\} 
\,. 
\label{g-fmass}
\end{eqnarray} 
The couplings $z_3$ and $z_7$ are expected to be of the one-loop order, ${\cal O}(N_c/(4\pi)^2)$. 
The HLS gauge coupling $g \sim m_\rho/F_\pi$ from the Lagrangian Eq.(\ref{1-formLagrangian}), 
so that the correction terms in the second parenthesis of the second line in Eq.(\ref{g-fmass}) 
certainly contribute to the square of the mass as the ${\cal O}(p^4)$ correction 
$\sim  eB \cdot(\frac{m_\rho}{(4\pi F_\pi)})^2 \sim {\cal O}(p^4)$, 
based on the derivative/chiral expansion in Eq.(\ref{expansion-parameters}).

\begin{figure}[t] 
\centering
\includegraphics*[scale=0.6]{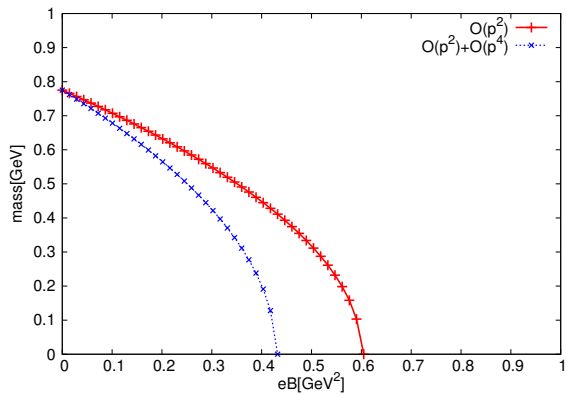}
\caption{ 
The plot of the effective mass for the charged $\rho$ meson 
in Eq.(\ref{g-fmass}) 
as a function of the external magnetic field $(eB)$, 
with the central value of the anomalous magnetic moment in Eq.(\ref{val:anomag}) 
and the experimental value of $m_\rho$, $m_\rho=0.775\, [{\rm GeV}]$~\cite{Agashe:2014kda} 
(dotted curve marked by ``x''). 
Also has been shown the effective mass without the anomalous magnetic moment, 
corresponding to Eq.(\ref{G:op2}) (denoted by the curve marked as ``+'').  
} 
\label{xyrhoO(p2)O(p4)p}
\end{figure}

We may fix the anomalous magnetic-moment term in Eq.(\ref{g-fmass}) by quoting 
the recent result from the lattice QCD~\cite{Luschevskaya:2015bea} 
to find 
\begin{eqnarray}
-g^2 z_3 +\frac{g^2z_7}{2}= 0.4 \pm 0.2
\,.  
\label{val:anomag}
\end{eqnarray}
(The amount of the ${\cal O}(p^4)$ coefficients $z_3 $ and $z_7$ fixed here 
is consistent with the expected order based on the derivative/chiral expansion.) 
In Fig.~\ref{xyrhoO(p2)O(p4)p}, we plot the effective mass in Eq.(\ref{g-fmass}) 
as a function of  the external magnetic field $(eB)$, 
with the central value of the anomalous magnetic moment in Eq.(\ref{val:anomag}) 
and the experimental value of $m_\rho$, $m_\rho=0.775\,[{\rm GeV}]$~\cite{Agashe:2014kda}. 
To see the effect of the anomalous magnetic moment, we have also shown 
the effective mass at the order of ${\cal O}(p^2)$ in Eq.(\ref{G:op2}) in the figure. 
We see from the figure 
that as the magnetic scale $(eB)$ grows,   
the rate of reduction for the effective mass gets enhanced by 
the anomalous magnetic moment. 
At $eB=0.3[{\rm GeV}^2]$, for instance, 
 the effective mass in Eq.(\ref{G:op2}) is estimated to be 
$m_\rho(eB=0.3)\simeq 0.55 \, [{\rm GeV}]$, 
which decreases by about 20\% to be  
$\simeq 0.43 \, [{\rm GeV}]$ due to the anomalous magnetic moment in Eq.(\ref{g-fmass}).


\subsection{Neutral Vector Meson Masses }\label{B}

We now turn to the neutral vector mesons. 
At the leading order of the derivative/chiral 
expansion, 
their masses cannot, of course, be affected by the magnetic field 
as clearly seen from the Lagrangian Eq.(\ref{frhoLagrangian}). 
Even at the higher order as in Eq.(\ref{z3z7}), 
it seems still true so one might conclude that the neutral vector mesons are 
free from the magnetic-scale dependence. 
However, it is actually not the case:  
the point to note is that all terms in Eqs.(\ref{frhoLagrangian}) and (\ref{z3z7}) preserve 
the intrinsic parity (IP) -- which is defined to be even when 
a particle has the parity $(-1)^{\rm spin}$, otherwise odd. 
In the HLS model, IP-odd terms can be incorporated at the order of ${\cal O}(p^4)$, 
which are HLS-gauge invariant (a la Wess-Zumino-Witten) terms giving rise to 
the anomalous vector meson decay processes such as 
$\omega \to \pi^0 \gamma$ and $\rho^0 \to \pi^0 \gamma$~\cite{Fujiwara:1984mp,Harada:2003jx}. 
As will be definitely shown in this subsection, 
the neutral vector meson masses, along the magnetic direction, 
can be affected by the magnetic field through the IP-odd terms 
to get modified by terms 
on the order of $(eB)^2$. 
The result presented here gives the prediction to be tested by future lattice simulations.

\begin{figure}[t]
\centering
\includegraphics[width=16cm]{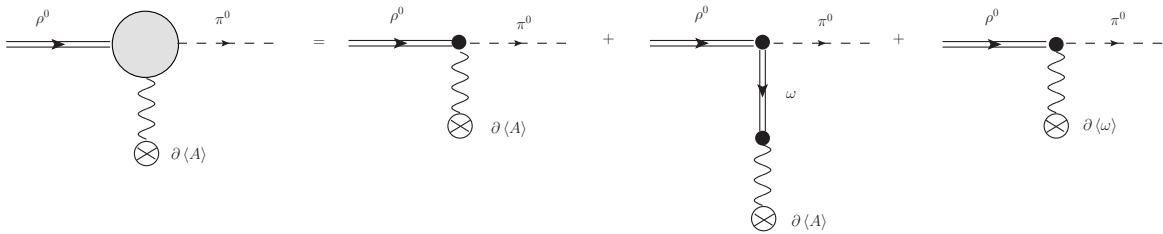}
\vspace*{-6cm}
\caption{
The Feynman graphs corresponding to the effective vertex function for the 
$\rho^0$-$\pi^0$-$\langle A \rangle$. 
The crossed circle denotes the background fields $\langle A \rangle$ 
and $\langle \omega \rangle=e/(3g') \langle A \rangle$ 
multiplied by derivative.  
}
\label{rhopisum}
\end{figure}

The IP-odd sector involving the vector mesons is constructed from 
four terms written in the HLS-gauge invariant way~\cite{Harada:2003jx}. 
Among them, one finds that the following two terms are relevant to the present study: 
\begin{eqnarray}
{\cal L}_{c}&=&
-\frac{N_c}{16\pi^2}c_3\epsilon^{\mu\nu\lambda\sigma}{\rm tr}[\{\hat\alpha_{\perp\mu},\hat\alpha_{\parallel\nu}\}V_{\lambda\sigma}]\nonumber\\
&&-\frac{N_c}{16\pi^2}c_4\epsilon^{\mu\nu\lambda\sigma}{\rm tr}[\{\hat\alpha_{\perp\mu},\hat\alpha_{\parallel\nu}\}\hat{\cal V}_{\lambda\sigma}]
\label{c3c4term}
\end{eqnarray}
with $N_c=3$.  
When the magnetic field is frozen as in Eq.(\ref{symmetry}), 
these IP-odd terms turn out to 
induce the $\rho^0$-$\pi^0$-$\omega$ mixing 
at the order of ${\cal O}(eB)$: 
combining the $c_3$ and $c_4$ terms 
with the $\rho^0$ - $\langle A \rangle$ 
and $\omega$ - $\langle A \rangle$ mixing arising from the IP-even sector in Eq.(\ref{frhoLagrangian}), 
we construct the effective $\rho^0$ - $\pi^0$ - $\langle A \rangle$ 
and $\omega$ - $\pi^0$ - $\langle A \rangle$ vertex functions.  
Figure~\ref{rhopisum} depicts the Feynman graphs for one of them. 
The resultant effective vertex functions can be cast into the 
following operator form: 
\begin{eqnarray}
{\cal L}_{{\rm eff}}= 
\epsilon^{xy\lambda\sigma} eB \left[ 
g_{\rho\pi\gamma} \, \pi^0\d_\lambda\tilde{\rho}^0_\sigma
+g_{\omega\pi\gamma} \, \pi^0\d_\lambda\tilde{\omega}_\sigma
\right] 
\,, 
\end{eqnarray}
where 
\begin{eqnarray}
g_{\rho\pi\gamma}&=& 
- \frac{N_cg}{24\pi^2F_\pi} \left( -c_3+\frac{c_3-c_4}{2}  \right) 
\,, \nonumber \\
g_{\omega\pi\gamma}&=& 
- \frac{N_cg'}{8\pi^2F_\pi} \left(-c_3+\frac{c_3-c_4}{2}\right) 
\,. 
\label{ipmixing}
\end{eqnarray} 
The first term in the parentheses arises from a type of the first diagram in Fig.~\ref{rhopisum}, 
while the second term from the third diagram. The second diagram in Fig.~\ref{rhopisum}, 
involving the vector meson propagator, 
actually does not contribute at all, because of the constant magnetic field.   
Thus the IP-odd terms generate the vector meson mixing with the neutral pion.

Including the $\rho^0$-$\omega$-$\pi^0$ mixing,   
we write the $\rho^0$, $\omega$ and $\pi^0$ propagators in the matrix form:  
\begin{eqnarray}
&&
-\frac{1}{2} 
\left( 
\begin{array}{c} 
\tilde{\rho}^{0z}(-p) \\ 
\pi^0(-p) \\ 
\tilde{\omega}^z(-p) 
\end{array} 
\right)^T 
\begin{pmatrix}
m_\rho^2-p_\mu p^\mu&ig_{\rho\pi\gamma}p^teB&0\\
-ig_{\rho\pi\gamma}p^teB&m_\pi^2-p_\mu p^\mu&-ig_{\omega\pi\gamma}p^teB\\
0&ip^tg_{\omega\pi\gamma}eB&m_\omega^2-p_\mu p^\mu
\end{pmatrix}
\begin{pmatrix}
\tilde{\rho}^{0z}(p)\\
\pi^0(p)\\
\tilde{\omega}^z(p)
\end{pmatrix}
\,, \label{propa:matrix}
\end{eqnarray}
where the label $t$ attached on momentum $p$ denotes the 0th component in the momentum space.  
Defining the effective masses at ${\bm p}=0$ (i.e., $p^t \equiv$ mass)
and considering a weak magnetic field satisfying 
\begin{eqnarray}  
g_{\rho/\omega\pi\gamma} \, \frac{eB}{m_{\rho/\omega}} \ll 1 
\label{approx}
\end{eqnarray}
(See Eq.(\ref{ipmixing})), 
we can analytically calculate the mass eigenvalues from the propagator matrix 
to get~\footnote{Looking at the matrix form in Eq.(\ref{propa:matrix}) or 
the formulae for the mass eigenvalues in Eq.(\ref{mass:formula}), 
one should note that the $(eB)^2$ corrections do not come with 
the mass difference $(m_\omega^2 - m_\rho^2)$ in the denominators. 
This is due to the absence of the direct $\rho^0$ - $\omega$ mixing 
up to the ${\cal O}(p^4)$. 
If it were present at this order, 
the finite widths of the vector mesons should  
be significant, since the $\rho$ and $\omega$ are almost degenerate. 
\label{foot}}  
\begin{eqnarray}
m_{\rho^0}(eB)&=& 
m_\rho\Bigl[1+\frac{1}{2(1-m_{\pi^0}^2/m_\rho^2)}
\left(\frac{eBg_{\rho\pi\gamma}}{m_\rho}\right)^2+\cdots\Bigl]
\,, \nonumber \\
m_{\pi^0}(eB)&=& 
m_{\pi^0}\Bigl[1-\frac{1}{2(1-m_{\pi^0}^2/m_\rho^2)}
\left(\frac{eBg_{\rho\pi\gamma}}{m_\rho}\right)^2-\frac{1}{2(1-m_{\pi^0}^2/m_\omega^2)}\left(\frac{eBg_{\omega\pi\gamma}}{m_\omega}\right)^2+\cdots\Bigl]
\,, \nonumber \\
m_\omega(eB)&=& m_\omega\Bigl[1+\frac{1}{2(1-m_{\pi^0}^2/m_\omega^2)}
\left(\frac{eBg_{\omega\pi\gamma}}{m_\omega}\right)^2+\cdots\Bigl] 
\,, 
\label{mass:formula}
\end{eqnarray}
Thus, in the weak magnetic scale region as in Eq.(\ref{approx}), 
the $\rho^{0}$ and $\omega$ masses along the magnetic direction 
tend to increase, while the $\pi^{0}$ mass gets smaller.

\begin{figure}[htbp]
\centering
\includegraphics*[scale=0.6]{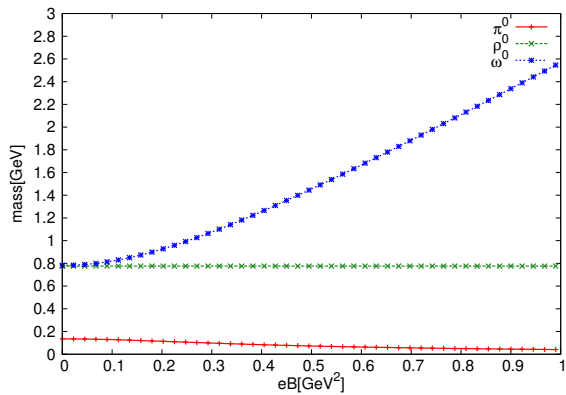}
\caption{
The plot of the effective masses for the $\pi^0$, 
$\rho^0$ and $\omega$ as a function of the constant magnetic field $(eB)$. 
Here $\rho^0$ and $\omega$ are polarized along the magnetic field ({\rm z}-direction).
}
\label{ipoddmassshift}
\end{figure}

Apart from the approximation in Eq.(\ref{approx}), 
we numerically solve the mixing matrix. 
To this end, 
we fix the values of the $g_{\rho\pi\gamma}$ and $g_{\omega\pi\gamma}$ 
couplings in Eq.(\ref{propa:matrix})
by using the experimental values of the decay widths 
$\Gamma(\rho^0\rightarrow\pi^0\gamma)$ and $\Gamma(\omega\rightarrow\pi^0\gamma)$: 
\begin{eqnarray}
\Gamma(\rho^0\rightarrow\pi^0\gamma) 
&=& 
\frac{\alpha}{24} |g_{\rho\pi\gamma}|^2 \Bigl(\frac{m_\rho^2-m_{\pi^0}^2}{m_\rho}\Bigl)^3 
\simeq 8.9 \times 10^{-5}[{\rm GeV}]
\,, \nonumber \\
\Gamma(\omega\rightarrow\pi^0\gamma)&=&
\frac{\alpha}{24} |g_{\omega\pi\gamma}|^2 \Bigl(\frac{m_\omega^2-m_{\pi^0}^2}{m_\omega}\Bigl)^3
\simeq 7.0 \times 10^{-4}[{\rm GeV}]
\, ,  
\label{width}
\end{eqnarray}
where $\alpha = e^2/(4\pi)$ and 
the left hand sides are theoretical expressions obtained by the present model. 
Using the experimental values, $\alpha \simeq 1/137$ and $m_{\pi^0}=0.135[{\rm GeV}]$, 
$m_{\rho}=0.775[{\rm GeV}]$ and 
$m_{\omega}=0.783[{\rm GeV}]$~\cite{Agashe:2014kda},  
we have 
\begin{eqnarray}
|g_{\rho\pi\gamma}|   & \simeq 8.3 \times 10^{-1}[{\rm GeV^{-1}}] 
\,, \nonumber \\
|g_{\omega\pi\gamma}| & \simeq 2.3 \times 10^{0}[{\rm GeV}^{-1}]
\,. 
\end{eqnarray}
In Fig.~\ref{ipoddmassshift}, 
we plot the effective masses of 
the $\pi^0$, $\rho^0$ and $\omega$  
as a function of the external magnetic field $(eB)$.  
 As expected from the approximate formulae in Eq.(\ref{mass:formula}), 
the vector meson masses increase as the magnetic scale gets larger. 
Most prominently,  
the $\omega$ mass dramatically gets enhanced. 
For instance, the mass becomes larger  
by about $37\%$ when the $(eB)$ reaches the scale around 0.3 ${\rm GeV}^2$.  
  This significant enhancement is testable 
in future lattice simulations.

\section{Summary and Discussion}
\label{summary}

In this paper, we discussed the magnetic responses of vector meson masses 
based on the hidden local symmetry model in a constant magnetic field.   
We employed the lightest two-flavor system including 
the pion, rho and omega mesons in the spectrum. 
The effective masses influenced under the magnetic field were  
evaluated in a way of 
the derivative/chiral expansion established in the hidden local symmetry model. 
At the leading order the g-factor of the charged rho meson is 
fixed to be 2, implying that the rho meson at this order looks as if it were a point-like spin-1 particle.  
Going beyond the leading order,  
we found anomalous magnetic interactions of the charged rho meson, 
involving the anomalous magnetic moment arising 
from ${\cal O}(p^4)$ terms.  
It was suggested that the charged rho mass can be vanishing 
up to the order of ${\cal O}(p^4)$. 
More remarkably,  
nontrivial magnetic-dependence of neutral mesons emerges 
to give rise to the significant mixing among neutral mesons 
through the intrinsic-parity odd term of ${\cal O}(p^4)$. 
We found the dramatic enhancement of the omega meson mass, 
which is testable in future lattice simulations.

Phenomenological consequences derived from 
the enhanced omega mass in magnetic field 
are anticipated to be seen through decay processes relevant to 
strong-magnetic field systems such as in magnetars. 
Inclusion of temperature into the present model is straightforward, 
so it would be possible to draw some implications 
to the magnetic catalysis or 
inversed one, and physic in heavy ion collisions. 
More explicit analysis on such exotic phenomenologies 
is to be pursued elsewhere. 
\\

Before closing the present paper, we discuss possible corrections to our findings from 
terms higher than ${\cal O}(p^4)$.  
\\

%

Based on the systematic expansion with respect to the derivative/chiral orders, 
established in the HLS model, 
in the present paper we have discussed magnetic responses of vector mesons. 
As noted in the end of Sec.~\ref{effective-masses}, 
one would say that the charged rho meson can be massless when one looks at 
Fig.~\ref{xyrhoO(p2)O(p4)p}, and 
one might have a question: could it be true even if higher order corrections 
are incorporated? 
First of all, we shall consider this point. 
\\

{\it ${\cal O}(eB)^2$ corrections to 
charged $\rho$ mass transverse to magnetic field  } ---  
The charged rho mass transverse to the magnetic field has been evaluated 
including terms up to ${\cal O}(p^4)$ in the derivative/chiral expansion of the 
HLS model. Those terms come in the mass as the correction of ${\cal O}(eB)$, 
i.e., the magnetic moment term (See Eq.(\ref{g-fmass})). 
Figure \ref{xyrhoO(p2)O(p4)p} implies that 
the $\rho$ meson can be massless at around $eB \simeq 0.3 - 0.4\, [{\rm GeV^2}]$. 
However, as noted in the literature~\cite{Hidaka:2012mz},  
the vanishing charged rho mass in the magnetic field might  
contradict with the Vafa-Witten theorem~\cite{Vafa:1983tf} 
and also the QCD inequalities~\cite{Weingarten:1983uj,Witten:1983ut,Nussinov:1983hb,Espriu:1984mq,Nussinov:1999sx}. 

Here we shall attempt to include 
corrections of the order of ${\cal O}(eB)^2$, 
which would arise from ${\cal O}(p^6)$ terms.
One can easily write down ${\cal O}(p^6)$ terms in a manner 
invariant under the chiral/HLS, parity and charge conjugations. 
The terms relevant to the charged rho mass then involve the following operators~\footnote{
Other ${\cal O}(p^6)$ terms made of product of two traces, 
such as ${\rm tr}[\hat{\cal V}_{\mu\nu}^2] {\rm tr}[\hat{\alpha}_{||\sigma}^2]$, 
can be incorporated there, which would, however, be suppressed by a factor of $(1/N_c)$ 
compared to terms with coefficients $d_{1,2,3,4,5}$ in Eq.(\ref{ds}).  
}: 
\begin{eqnarray}
{\cal L}_{(6)}
&=&
\frac{1}{(4\pi F_\pi)^2}
\bigl\{
d_1{\rm tr}[\hat{\cal V}^{\mu\nu}\hat{\cal V}_{\mu\nu}\hat\alpha_{\parallel\sigma}\hat\alpha_{\parallel}^\sigma]+
d_2{\rm tr}[\hat{\cal V}^{\mu\nu}\hat{\cal V}_{\mu\sigma}\hat\alpha_{\parallel\nu}\hat\alpha_{\parallel}^\sigma]+
d_3{\rm tr}[\hat{\cal V}^{\mu\nu}\hat{\cal V}_{\mu\sigma}\hat\alpha_{\parallel}^{\sigma}\hat\alpha_{\parallel\nu}] 
\nonumber \\ 
&&
-d_4\bigl(
{\rm tr}[\hat{\cal V}^{\mu\nu}\hat\alpha_{\parallel\nu}\hat{\cal V}_{\mu\sigma}\hat\alpha_{\parallel}^\sigma]+
{\rm tr}[\hat{\cal V}^{\mu\nu}\hat\alpha_{\parallel}^{\sigma}\hat{\cal V}_{\mu\sigma}\hat\alpha_{\parallel\nu}]
\bigl)-
d_5{\rm tr}[\hat{\cal V}^{\mu\nu}\hat\alpha_{\parallel\sigma}\hat{\cal V}_{\mu\nu}\hat\alpha_{\parallel}^\sigma]\bigl\}
\,, \label{ds}
\end{eqnarray}
where $d_{1,2,3,4,5}$ are arbitrary coefficients, expected to be on the order of 
${\cal O}(N_c/(4\pi)^2) = {\cal O}(10^{-2})$ from the scheme in terms of the 
systematic derivative/chiral expansion. 
Taking these ${\cal O}(p^6)$ terms into account, 
one finds that 
the charged rho mass transverse to the magnetic field (in the LLL) is modified from Eq.(\ref{g-fmass}) to be~\footnote{
 Other $B$-dependent terms to the rho mass formula in Eq.(\ref{rhop4})  would arise from charged pion loops  
as finite part corrections like $\sim 1/(4 \pi)^2 \log (eB/m_\pi^2)$. 
This term can be absorbed into redefinition of 
the HLS couplings, such as $g$, to give a shift to the ${\cal O}(p^6)$ term in Eq.(\ref{rhop4}) as $\sim (eB) \log(eB)$, 
which is, however, negligibly small compared to the power correction term $\sim (eB)^2$.   
} 
\begin{eqnarray}
\Bigg\{ m_\rho^2-eB \Bigg\} 
+ 
\Bigg\{  \left(z_3 - \frac{z_7}{2} \right) g^2 \, eB \Bigg\} 
 + 
\Bigg\{ \frac{d}{4(4\pi F_\pi)^2} g^2 (eB)^2 \Bigg\} 
\,, 
\label{rhop4}
\end{eqnarray}
with $d = 2 d_1+d_2+d_3+2d_4+2d_5$.

\begin{figure}[t]
\centering
\includegraphics*[scale=0.6]{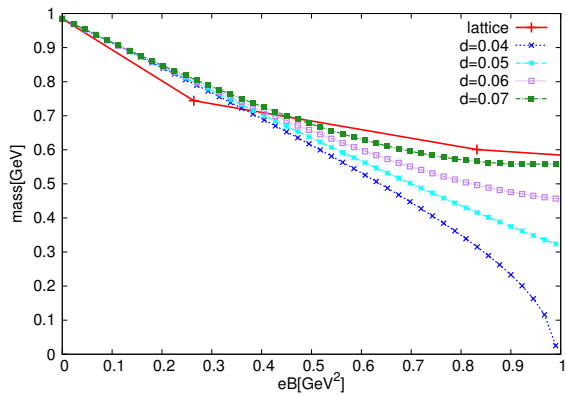}
\caption{
The charged rho meson mass transverse to the magnetic field as a function of $(eB)$ 
in Eq.(\ref{rhop4}) with the ${\cal O}(p^6)$ corrections included, 
taking the ${\cal O}(p^6)$ parameter $d$ as $d=0.04,0.05,0.06,0.07$.   
The curves have been compared with the current lattice data~\cite{Hidaka:2012mz}. 
In the plots the rho mass in the vacuum has been set to the value estimated in the lattice simulation. 
}
\label{Op6crho}
\end{figure}

To see how corrections from the ${\cal O}(p^6)$ terms can be effective, 
in Fig.~\ref{Op6crho} we attempt to plot the effective mass in Eq.(\ref{rhop4}) as a function of $(eB)$ 
taking the coefficient $d=0.04,0.05,0.06,0.07$, 
where the rho mass in the vacuum has been set to the value estimated in the lattice simulation~\cite{Hidaka:2012mz}, 
$m_\rho = 985$ MeV,  
and we have used $F_\pi = 92$ MeV~\cite{Agashe:2014kda}. 
The HLS gauge coupling $g$ has been estimated by assuming the vector meson dominance for 
the pion electromagnetic form factor~\cite{Harada:2003jx} 
as $g \simeq 7.55$ with use of the values of $m_\rho$ and $F_\pi$ as above.   
We see from the figure that  
 the ${\cal O}(p^6)$ correction can kick up
the effective mass from the massless limit point 
(around $eB \simeq 0.3 - 0.4\, {\rm GeV}^2$).
Similar observation 
in a region of strong magnetic scale $\gtrsim 1\,{\rm GeV}^2$    
has been made in a quark model approach~\cite{Taya:2014nha}, 
where it is proposed that the increase of effective mass for the rho meson 
can be understood by magnetic moments of constituent quarks, not in terms of 
hadrons. 
In Fig.~\ref{Op6crho} 
such a degree of freedom (D.O.F.) of constituent quarks might be mimicked  
by higher derivative term corrections.  
Incidentally, in Fig.~\ref{Op6crho} comparison  with the lattice data points~\cite{Hidaka:2012mz},  
that is based on the quenched QCD with no quark loop effects,  
has also been made. 
\\

{\it ${\cal O}(eB)^2$ corrections to $\omega$ mass along the magnetic field} -- 
When the ${\cal O}(p^6)$ corrections would be relevant as seen from Fig.~\ref{Op6crho}, 
one might suspect that the significant enhancement of the $\omega$ meson mass   
in Fig.~\ref{ipoddmassshift}   
could be washed out by higher order corrections. 
However, it is not the case: 
the $\omega$ meson mass, parallel to the magnetic field, 
would get contributions from terms of ${\cal O}(p^6)$ (i.e. ${\cal O}(eB)^2$) 
in the form constructed from two traces like   
\begin{equation} 
 {\rm tr} [\hat{\cal V}_{\mu\nu}^2 \hat{\alpha}_{|| \sigma}] 
{\rm tr}[\hat{\alpha}_{||}^\sigma] 
\,. \label{trtr}
\end{equation} 
Note that this class of contributions is potentially suppressed in terms of 
the $1/N_c$ expansion (1/3 suppression, in the real-life QCD), 
in addition to the suppression by extra loop factor, 
compared to IP-odd terms of the leading order of 
${\cal O}(N_c)$ (See Eq.(\ref{c3c4term})). 
We have numerically checked that the ${\cal O}(p^6)$ corrections 
are indeed tiny enough to keep the result in Fig.~\ref{ipoddmassshift} 
intact, up to a moderate magnetic scale, say, $eB \simeq 0.3\, {\rm GeV}^2$. 
Thus, ${\cal O}(p^6)$ (${\cal O}(eB)^2$) corrections to the $\omega$ meson mass 
can safely be neglected.

The ${\cal O}(p^6)$ operator as in Eq.(\ref{trtr}) can actually generate 
the direct $\rho^0$-$\omega$ mixing. 
Even if the $\rho^0$-$\omega$ mixing is absent at the order of ${\cal O}(p^4)$, 
as noted in footnote~\ref{foot}, 
one might naively think that the higher order corrections 
would spoil the argument in the enhancement on the $\omega$ meson mass 
in Sec.~\ref{effective-masses}. 
However, it is again not the case: 
by the same argument as above, 
the ${\cal O}(p^6)$ contribution turns out to be sufficiently 
suppressed in magnitude by the extra loop factor and $1/N_c(\sim 1/3)$, 
compared to terms of ${\cal O}(p^4)$. 
Thus the $\rho^0$ - $\omega$ mixing will not be sizable to be negligible 
in the mixing structure as in Eq.(\ref{propa:matrix}),     
even considering the higher order terms. 
In the end, the significant development on 
the $\omega$ mass predicted in 
Sec.~\ref{effective-masses} is totally intact. 
\\

In closing, 
as has been addressed so far, 
the result including ${\cal O}(p^4)$ corrections shown in 
Figs.~\ref{xyrhoO(p2)O(p4)p} and \ref{ipoddmassshift}  
would be reliable up to some moderate magnetic scale 
$eB \simeq 0.3 - 0.4 \, {\rm GeV}^2$~\footnote{In this respect, 
the neutral pion in Fig.~\ref{ipoddmassshift} 
cannot be massless, although 
the effective mass tends to decrease monotonically 
as the magnetic scale gets larger.}.    
Going further to a larger magnetic scale region 
would need to incorporate higher order terms to 
such as  
corrections of ${\cal O}(p^6)$ (i.e. ${\cal O}(eB)^2$), 
in order that one appropriately discusses the magnetic 
response of vector 
meson masses.

\acknowledgments
The authors would like to thank Yoshimasa Hidaka for
enlightening discussions 
and  Masayasu Harada and Hiroki Nishihara  for useful comments. 
This work was supported in part by 
the JSPS Grant-in-Aid for Young Scientists (B) \#15K17645 (S.M.).

\appendix
\section{The Landau Quantization of the Vector Meson Mass}
\label{Landau}

We derive the equations of motion for $\phi$ and 
$\Phi$ fields from the Lagrangian Eq.(\ref{phiLagrangian})  
\begin{eqnarray}
&&M_1\phi-\frac{1}{2}H_1H_2\phi+\frac{1}{2}H_1H_1\Phi=0\label{phimotion}
\,, \\
&&M_2\Phi-\frac{1}{2}H_2H_1\Phi+\frac{1}{2}H_2H_2\phi=0
\,, 
\end{eqnarray}
where 
\begin{eqnarray} 
M_1  &=& m_\rho^2-eB+\d_t^2-\d_z^2
\,, \nonumber \\  
M_2 &=& m_\rho^2+eB+\d_t^2-\d_z^2
\,,  \nonumber \\ 
H_1 &=& D_x-iD_y 
\,, \nonumber \\   
H_2 &=& D_x+iD_y 
\,.
\end{eqnarray}
We then multiply by $H_1H_1$ on the left hand side, 
and use the commutation relation $[H_1, H_2]=2eB$ to get 
\begin{eqnarray}
\left( 
M_2-\frac{1}{2}H_1H_2-eB 
\right) 
H_1H_1\Phi+\frac{1}{2}H_1H_1H_2H_2\phi=0
\,. 
\end{eqnarray}
Using Eq.~(\ref{phimotion}) to replace $\Phi$ with $\phi$, 
we have 
\begin{eqnarray}
(m_\rho^2+\d_t^2-\d_z^2)(m_\rho^2+\d_t^2-\d_z^2-H_1H_2 -eB)\phi=0 
\,. 
\end{eqnarray}
Note that $H_1$ and $H_2$ form the creation $(a^\dag)$ and annihilation ($a$) operators 
for the Landau quantization:   
\begin{eqnarray}
a&=&\frac{1}{\sqrt{eB}}(\d_{\overline{Z}}+\frac{eB}{2}Z) 
\,, \nonumber\\
a^\dagger&=&\frac{1}{\sqrt{eB}}(-\d_{Z}+\frac{eB}{2}\overline{Z}) 
\,, \nonumber\\
\end{eqnarray}
where 
\begin{eqnarray} 
Z &=& (x+iy)/\sqrt{2} \,, \nonumber \\  
\d_{Z} &=& \frac{(\d_x-i\d_y)}{\sqrt{2}} 
\,, \nonumber \\ 
&& [a,a^\dagger] = 1 
\,. 
\end{eqnarray} 
Since the eigenvalue of $(a^\dagger a)$ is integer $n$, 
the energy (mass) of the $\rho$ meson is thus given by  
\begin{eqnarray}
E(eB)_n=\sqrt{(p^z)^2+m_\rho^2+(2n-1)eB} 
\,. 
\end{eqnarray}


\end{document}